# Size and location of radish chromosome regions carrying the fertility restorer *Rfk1* gene in spring turnip rape

Tarja Niemelä, Mervi Seppänen, Farah Badakshi, Veli-Matti Rokka and J.S.(Pat) Heslop-Harrison

T. Niemelä, M. Seppänen
Department of Agriculture, University of Helsinki,
PO Box 27, FI-00014 University of Helsinki, Finland
e-mail: tarja.niemela@helsinki.fi

V-M. Rokka
MTT Agrifood Research Finland, Biotechnology and Food Research,
FI-31600 Jokioinen, Finland

J.S. Heslop-Harrison, Farah Badakshi
Department of Biology, University of Leicester,
LE1 7RH, Leicester, UK

**Abbreviations**

BAC-FISH bacterial artificial chromosome – fluorescence *in situ* hybridization

CMS cytoplasmic male sterility

GISH genomic *in situ* hybridization

Rf restore fertility

**Running title**     Size and location of *Rfk1* gene

**Abstract**

In spring turnip rape (*Brassica rapa* L. spp. *oleifera*) the most promising F1 hybrid system would be the Ogu-INRA CMS/Rf system. A Kosena fertility restorer gene *Rfk1*, homologue of the Ogura restorer gene *Rfo*, was successfully transferred from oilseed rape into turnip rape and that restored the fertility in female lines carrying Ogura cms. The trait was, however, unstable in subsequent generations. The physical localization of the radish chromosomal region carrying the *Rfk1* gene was investigated using GISH (genomic *in situ* hybridization) and BAC-FISH (bacterial artificial chromosome – fluorescence *in situ* hybridization) methods. The metaphase chromosomes were hybridized using radish DNA as the genomic probe and BAC64 probe, which is linked with *Rfo* gene. Both probes showed a signal in the chromosome spreads of the restorer line 4021-2 Rfk of turnip rape but not in the negative control line 4021B. The GISH analyses clearly showed that the turnip rape restorer plants were either monosomic ($2n=2x=20+1R$) or disomic ($2n=2x=20+2R$) addition lines with one or two copies of a single alien chromosome region originating from radish. In the BAC-FISH analysis, double dot signals were detected in sub-terminal parts of the radish chromosome arms showing that the fertility restorer gene *Rfk1* was located in this additional radish chromosome. Detected disomic addition lines were found to be unstable for turnip rape hybrid production. Using the BAC-FISH analysis, weak signals were sometimes visible in two chromosomes of turnip rape and a homologous region of *Rfk1* in chromosome 9 of the *B. rapa* A genome was verified with BLAST analysis. In the future this homologous area in A genome could be substituted with radish chromosome area carrying the *Rfk1* gene.

**Introduction**

Spring turnip rape (*Brassica rapa* L. spp. *oleifera*, 2n=2x=20, genome constitution AA) is the major oilseed crop cultivated for the production of vegetable oil and animal feed protein in many northern areas including Finland, parts of Canada and Northern India. Compared to oilseed rape (*Brassica napus* L., 2n=4x=38), which has a higher yield potential, turnip rape is early maturing and therefore shows better yield stability in Northern climates. Most of the cultivated varieties of turnip rape are open-pollinated, despite the potential of hybrid breeding exploiting heterosis effects, which could increase the seed yield (Niemelä et al. 2006). One of the main systems for F1 seed production in the genus *Brassica* is based on utilization of cytoplasmic male sterility (CMS) and fertility restoring genes (Rf). The CMS/Rf type of hybrid system is used commercially in many important agricultural crops, dominating production in maize (*Zea mays*) and sunflower (*Helianthus annuus*), with more limited and special uses in rice (*Oryza sativa*), cotton (*Gossypium hirsutum*) and oilseed rape (*B. napus*). In turnip rape there is no fully functional CMS/Rf hybrid system in commercial use. However, one of the most promising of the known hybrid systems is the Ogu-INRA CMS/Rf (Niemelä et al. 2010).

The Ogu-INRA CMS/Rf system was originally transferred from Japanese radish (*Raphanus sativus*; 2n=2x=18, genome RR) to oilseed rape (*Brassica napus*; genome AACC) (Bannerot et al. 1974; Bannerot et al. 1977; Heyn 1976). Nowadays it is one of the most commonly used systems for F1 hybrid production in oilseed rape breeding programmes. The Ogura cms-associated gene, *orf138*, induces abnormal flower development that prevents the production of functional pollen (Bonhomme et al. 1992; Krishnasamy and Makarof 1994). This male sterile line is called A-line in the hybrid seed production and the character is maintained using B-lines that have the normal cytoplasm without the sterility inducing traits. Those lines which restore fertility trait are called R-lines. In oilseed rape hybrids with the Ogura system, restorer lines carry one dominant nuclear gene, *Rfo* (Bellaoui et al.

1999). In oilseed rape, the *Rfo* restorer gene has been optimized to not contain flanking chromosomal sequences from the original radish introgression (SOFIPROTEOL European Patent No EP1382612 and PCT No WO02088179 (1). European Patent No EP1556495 and PCT No WO04039988 (2) – retrieved from http://www.pbltechnology.com/cms.php?pageid=322 20.10.2011).

Previously, Ogura cms was transferred from oilseed rape into turnip rape (Sovero 1987; Delourme et al. 1994) resulting in production of stable male sterile lines for use in turnip rape hybrid production. The transfer of the fertility restorer gene (*Rfo*) from oilseed rape into turnip rape has however been unsuccessful (F. Stoenescu, personal communication, Zeneca Seeds, Winnipeg, Canada, 1995). In recent studies, it was confirmed that the *Rfo* gene is introgressed into the C genome in oilseed rape (Hu et al. 2008; Feng et al. 2009). This location of the *Rfo* gene in the C genome may be one of the reasons its transfer into the turnip rape A genome has not yet succeeded. Therefore, to establish a specific hybrid system for turnip rape, the Kosena fertility restorer gene (*Rfk1*), originating from radish and homologue of the Ogura fertility restorer gene (*Rfo*) (Brown et al. 2003), was transferred from oilseed rape into turnip rape through interspecific crosses followed by traditional backcrossing (Niemelä et al. 2010). In contrast to the *Rfo*, the *Rfk1* gene was supposedly not integrated into the C genome of the oilseed rape breeding lines, selected for our turnip rape hybrid breeding program. During the course of the breeding program, it was observed that the *Rfk1* gene was able to restore the fertility of turnip rape with Ogura cms, but the trait was unstable in the turnip rape genome (Niemelä et al. 2010). After subsequent selection and interpollination of homozygous plants, the progeny also had some heterozygous and male sterile plants. To advance the lines for improvement of the fertility-restoring turnip rape male lines, it would be helpful to detect and measure the amount of introgressed radish genome and its location in the turnip rape A genome.

 

The physical localization of the radish introgression carrying the *Rfk1* gene in turnip rape genome can be investigated by fluorescence *in situ* hybridization (FISH) and genomic *in situ* hybridization (GISH). This approach has successfully been used to detect different genomes in interspecific hybrids as well as for locating introgressed genomes, alien chromosomes, or chromosomal segments in another genomic background (Schwarzacher et al. 1992). The recent progress in *Brassica* genome sequencing projects have provided useful sequence information for cytogenetic studies (review: Heslop-Harrison and Schwarzacher 2011) and *Brassica* bacterial artificial chromosome (BAC) clones have been used for hybridization as FISH probes to physically localize specific sequences on *Brassica* chromosomes (Howell et al. 2005, 2008; Nicolas et al. 2007, 2008; Feng et al. 2009; Kim et al. 2009; Szadkowski et al. 2010, 2011; Xiong and Pires 2010; Xiong et al. 2011). As in most plant species, transposable elements are abundant in the *Brassica* genome, and some of these are genome specific (Alix et al. 2005, 2008; Lim et al. 2007); there are also tandemly repeated DNA motifs which show specificity to some chromosomes and genomes (Harrison and Heslop-Harrison 1995). Along with other genome-specific sequences, this means that total genomic DNA can be used to distinguish genomes (Snowdon et al. 1997), but because of the relatively small chromosomes and limited evolutionary divergence of the A, B and C genomes, it is difficult to confirm the origin of all chromosomes throughout their length. However with the greater evolutionary distance between genera, *Brassica* and *Raphanus*, the separation is robust. Total genomic DNA of radish has been used as a probe to distinguish the radish R genome from A and C genomes of *Brassicas* (Snowdon et al. 1997; Benabdelmouna et al. 2003; Chen and Wu 2008; Akaba et al. 2009). Budahn et al. (2008) also used a radish specific probe, pURsN, *in situ* hybridization, to identify radish chromosome additions in oilseed rape.

The aim of the current study was to localize physically the radish *Rfk1* gene, the chromosome constitution, and the putative flanking region of radish chromatin in the turnip rape genomic background using GISH and BAC-FISH. Localization of the restorer gene aimed to understand the nature of its instability in the A genome and to suggest tools for selection and breeding towards the functional hybrid system for turnip rape.

**Material and methods**

Plant material

The breeding line, 4021-2 Rfk, of spring turnip rape (*Brassica rapa*) was selected for the chromosome preparations. Details of the breeding work of this restorer line 4021-2 Rfk has been described previously (Niemelä et al. 2010). The open pollinated turnip rape (*Brassica rapa*) line of Finnish origin, 4021B (AA, 2n=20), was used as a negative control for chromosome preparations. Both breeding lines had the same genetic background except the 4021-2 Rfk was carrying the Kosena *Rfk1* restorer gene. The 4021-2 Rfk was produced through traditional backcrosses, where the spring oilseed rape (*Brassica napus*) breeding line RfA4 (Plantech Research Institute Japan) having the Kosena *Rfk1* gene was used as a donor parent. The homozygous (Rfk1, Rfk1) plants were selected from BC6F4 progeny before flowering stage using TaqMan qPCR (Niemelä et al. 2010) and they were cross pollinated to form the fertility restoring 4021-2 Rfk line for the present study.

Chromosome preparation

Chromosome preparations were made from the root tips of turnip rape lines 4021B and 4021-2 Rfk using standard techniques (Schwarzacher and Heslop-Harrison 2000). In brief, seedling root tips were incubated in 2 mM 8-hydroxyquinoline for 3 h before fixation in fresh 3:1; ethanol : acetic acid. After storage and rinses, roots were digested in an enzyme solution [0.1% (w/v) cytohelicase (Sigma-Aldrich, Steinheim, Germany), 0.1% (w/v) cellulase Onozuka RS (Sigma-Aldrich, Steinheim, Germany), 0.1% Pectolyase Y23 (Sigma-Aldrich, Steinheim, Germany) in 10 mM citrate buffer, pH 4.8 for 90 min at room temperature, and then squashed in 60% (v/v) acetic acid. After freezing, coverslip removal and dehydration through an alcohol series, slides were selected under phase-contrast or after staining with DAPI (4,6-diamidino-2-phenylindole, Sigma), dehydrated, dried and stored at -20°C until hybridization.

Probe labelling

For the genomic probe, plant DNA was extracted from greenhouse-grown young leaves of radish 'Daikon' using DNeasy Plant Maxi Kits (Qiagen). DNA was sonicated to fragments of about 500 bp and labelled by random priming with biotin-11-dUTP (Roche) and digoxigenin-11-dUTP (Roche). For the BAC clone probe, BAC64 (Desloire et al. 2003) from Genoplante-Valor, was kindly provided by INRA-CNRGV (Centre National de Ressources Genomiques Vegetales, Castanet-Tolosan, France); the fertility restoration locus *Rfo* (homologue to *Rfk1*) is in *R.sativus* BAC64 (contig 127 kb, accession number AJ550021). BAC DNA was isolated with a NucleoBond Xtra Midi kits (Macherey-Nagel) and PCR using the primer pair (forward primer, 5'-TCATCCCCCAAATGATAGAT-3'; reverse primer, 5'-GAAGCTGCAAAGTGGGTTTC-3') designed for the *Rfk1* gene was carried out to verify the BAC64 identity. BAC DNA was sonicated to fragments <1kb, and labelled with biotin-11-dUTP or digoxigenin-11-dUTP using the Invitrogen BioPrime CGH labelling kit. Two ribosomal DNA probes, 5S and 45S, were labelled with Alexa-647-dUTP (Invitrogen).

*In situ* hybridization and signal detection

*In situ* hybridization was performed according to Schwarzacher and Heslop-Harrison (2000) and Schwarzacher (2008) with slight modifications. Up to three different probes were used in each hybridization, labelled with biotin-11-dUTP, digoxigenin-11-dUTP or Alexa-647-dUTP. Slides with chromosome spreads were re-fixed in ethanol:acetic acid 3:1, treated with RNase (100 µg/ml) solution and then with pepsin (5 µg/ml in 0.01 M HCl) to remove cytoplasm observed surrounding the mitotic chromosome spreads. Preparations were fixed with paraformaldehyde, dehydrated in an ethanol series and air dried. The hybridization mixture consisted of 40% formamide, 2xSSC, 10% dextran sulphate, 1µg of salmon sperm DNA, 0.125 mM EDTA, 0.125% SDS and 1-4 µl (25 to 60 ng) of each labelled probe with the final volume of 40-42 µl of mixture for each slide. The hybridization mixture was denatured at 85°C for 10 min, cooled on ice for 10 min and then applied to the slides. The slides with the chromosomes and probes were then denatured at 75°C for 7 min and hybridized at 37°C for 16 h using a modified thermal cycler. The post-hybridization washes were carried out following with a low-stringency wash using 0.1xSSC without formamide at 42°C. The hybridization sites were detected using a mix of anti-digoxigenin conjugated to FITC (Roche) and streptavidin conjugated to Alexa 594 (Invitrogen). The preparations were counterstained with DAPI and mounted in AF1 medium (Citifluor, London, UK). Slides were examined with an epifluorescence Zeiss Axiophot microscope and images were captured with a ProgRes C12 cooled CCD camera. Images were processed using Adobe

Photoshop CS4 using only functions including contrast and brightness adjustment that affect the whole area of the image equally. For the *in situ* hybridization 4 out of 20 restorer line 4021-2 Rfk preparations and 2 out of 8 control line 4021B preparations were selected. In each *in situ* hybridization, 3 to 10 metaphase chromosome sets were studied.

**Results**

Both genomic and BAC clone probes used in this study were hybridized with the chromosome preparations of the restorer line 4021-2 Rfk. Genomic *in situ* hybridization with *Raphanus* DNA revealed that the fertility-restoring turnip rape line 4021-2 Rfk plants were either monosomic (2n=2x=20+1R) or disomic (2n=2x=20+2R) addition lines with one copy or two copies of an alien chromosome originating from *Raphanus* (Japanese radish). In the control turnip rape line 4021B, without the restorer trait, no strong signal was detected in any of the mitotic chromosome spreads studied.

*In situ* hybridization with the BAC64 clone carrying the fertility restoration locus *Rfo* resulted in strong double dots of hybridization signal, one on each sister chromatid (Figs. 1a, 1c, 1f), on the radish chromosome identified by GISH (Figs. 1b, 1d, 1e). Thus the fertility restoring *Rfk1* gene, homologue to *Rfo,* was located on this chromosome. The two signals of BAC64 were more specifically located in the sub-terminal parts (Figs. 1a, 1b) of the radish chromosome arm. With the image of control line, 4021B, no strong signal was seen.

In addition to the signals detected in radish chromosomes as a result of hybridization with BAC64 probe, two pairs of weaker signals (Figs. 1a-1d, 1g) were also sometimes visible on two chromosomes of the turnip rape A genome, indicating the location of a homoeologous region to *Rfk1* on the A genome. A BLAST analysis comparing the full length sequence of *R. sativus* BAC64 clone (AJ550021.2) with the whole genome of *B.rapa* subsp. *pekinensis* (The *Brassica rapa* Genome Sequencing Project Consortium 2011) identified a *B. rapa* subsp. *pekinensis* BAC clone KBrB025K04 (AC189288.2) with high homology. The whole sequence coverage between BAC64 clone and BAC KBrB025K04 clone was 45%. The BAC KBrB025K04 is situated on the largest chromosome, linkage group A09 (6.10.2011) and it carries the fertility restorer gene (*Rf*) (Pentatricopeptide repeat-containing protein, fertility restorer B) (KBrB025K04CG0180), which has 90% homology with *R. sativus ppr-B* gene (*Rfo/Rfk1* gene) situated in BAC64 (88044-88063, 88191-90235). This *B.rapa* fertility restorer gene *B* (KBrB025K04CG0180) is homologue to *R.sativus ppr-B* gene (Mora et al. 2010), but is unable to restore fertility. The radish *Rfo* locus consists of three close related genes in tandem, named *ppr-A*, *ppr-B* and *ppr-C*, which the *ppr-B* has the fertility restoration activity (Brown et al. 2003; Desloire et al. 2003; Koizuka et al. 2003). The homology between *R.sativus* BAC64 clone and *B.rapa* linkage group A09 was visualized in a dot-plot matrix (Fig. 2), around 7.1Mb from the end of the 37.12Mb assembly, to show the homology of all three ppr genes in BAC64 (*ppr-A* gene situation: 80291-80313, 80454-82484; *ppr-B* gene situation - see above; *ppr-C* gene situation: 101320-102240, 102502-102609) with *B.rapa* linkage group A09 and the dot-plot analysis demonstrates that the region covering all the ppr genes has two copies in the *B.rapa* background.

## Discussion

The results define the physical position of a chromosome region containing a fertility restorer gene *Rfk1* located on an additional radish chromosome in the A genome of spring turnip rape (*Brassica rapa*). *Rfk1* is a valuable gene for breeding because it allows development of hybrid varieties for turnip rape. The genomic radish probe clearly hybridized to the chromosome/chromosomes of radish in the fertility restoring turnip rape line 4021-2 Rfk (Figs. 1b, 1d, 1e, 1g). The signals were strong and the whole chromosome was evenly labelled allowing characterization of the radish addition in the turnip rape genome.

The BAC carrying the *Rfk1* locus, BAC64, was used a robust probe for identifying the locus in *Brassica* and *Raphanus*, with little cross hybridization elsewhere in the genome. BAC clones frequently contain many repetitive sequences which are homologous in the target genome, requiring high stringencies and blocking in FISH (Kim et al. 2002; Schwarzacher 2008) to identify loci of the specific traits linked to that BAC clone. Similar clear BAC-FISH signal results were also found by Feng et al. (2009) when localizing the *Rfo* gene in oilseed rape genome using two different *R.sativus* BAC probes (G62 and B420) linked to the *Rfo* locus.

In the backcross progeny from *B.napus* to *B.rapa* (Niemelä et al. 2010), the segregation ratio followed 30:70 instead of the expected 50:50 ratio, perhaps due to location of the *Rfk1* gene as a segment of an extra region of radish chromosome in the turnip rape A genome. The GISH results showing both monosomic and disomic additions in the offspring, previously selected for homozygosity, demonstrate the instability of the addition chromosome (Niemelä et al. 2010). During two generations of selecting and intercrossing 100% homozygous plants (analysed by TaqMan qPCR) the offspring segregated to 90% homozygous and 10% hemizygous plants (data not shown). Budahn et al. (2008)

4reported that plants having disomic additions of radish chromosomes in rape-radish lines were expected to have high stability, and in some cases, disomic additions can be stable (Chevre et al. 1991). However, here the disomic addition of radish chromosome in the A genome of turnip rape was not stable, making them unsuitable for hybrid seed production. In several studies, instability of alien chromosomes has been found in hybrids obtained from intercrosses between different *Brassica* species (Chevre et al. 1991; Peterka et al. 2004; Wei et al. 2010). Ge and Li (2007) and Ge et al. (2009) have demonstrated that the epigenetic phenomenon known as nucleolar dominance plays a role in alien chromosome stability in *Brassica* species. The similarity or difference in between these components in parental material could affect to the chromosome structure and function during mitotic and meiotic divisions in hybrids. To achieve complete stability of the *Rfk1* gene, also exploited in practical breeding, the integration of the gene to the A genome chromosomes of turnip rape may be required. The *Rfo* gene, which is stably integrated in the oilseed rape C genome (Hu et al. 2008; Feng et al. 2009), is now used for hybrid seed production (Budar et al. 2004).

In *in situ* hybridization with BAC64 clone the 90% homology between the studied *R.sativus ppr-B* gene and *B.rapa* spp. pekinensis KBrB025K4CG0180 gene shows that these genomic regions are homologous between A and R genomes, which could increase the opportunity of transferring the fertility restoring trait from additional radish chromosome to turnip rape chromosome via homoeologous recombination. Tang et al. (2008) found also multicolour BAC-FISH very useful in identifying chromosomal regions between tomato (*Solanum lycopersicum*) and potato (*Solanum tuberosum*), which gives the information utilized in breeding techniques in introgressing genes from wild *Solanum* species into cultivated crops. The clear BAC-FISH signal of the radish chromosome as well as the weak BAC-FISH signal at the turnip rape chromosome (A09) was found from the sub-terminal region. According to Kim et al. (2002) and Wang et al. (2007) BAC hybridization shows that

Niemelä T, Seppänen M, Badakshi F, Rokka VM, Heslop-Harrison JS. 2012. Size and location of radish chromosome regions carrying the fertility restorer Rfk1 gene in spring turnip rape. Chromosome Research. Apr;20(3):353-361. http://dx.doi.org/10.1007/s10577-012-9280-5 Page 12 of 23.

13intergenomic introgressions often occur at distal parts of chromosomes.. In our case it would be desirable to have the homologous fertility restoring region in chromosome area with higher levels of recombination activity. Based on recent study of Shirasawa et al. (2011) *R.sativus* and *B.rapa* share large homologous genomic regions but the order or composition of these genomic segments do not correspond. This high genetic homology in between *R. sativus* and *B.rapa* would facilitate transferring the fertility restoring trait from radish to turnip rape, but the genetic information that regulates the homologous pairing during meiosis should be interrupted to favour recombination between nonhomologous A and R genome. The increasing knowledge of the mechanisms and genes, that control crossovers provide useful tools for plant breeders to promote homoelogous recombination in case of exploiting useful traits through interspecific crosses (Snowdon 2007; Nicolas et al. 2008; Wijnker and de Jong 2008). According recent studies in *B.napus* one possible way to increase homeologous recombination is manipulating the plant karyotype (Nicolas et al. 2009; Leflon et al. 2010). They found that special *PrBN* (*Pairing Regulator in B.napus*) gene regulates homeologous pairing in *B.napus* haploids and that depends on a plant's chromosomal composition. According these results they suppose that the increase in recombination could be due to change in ploidy level and it might work in more general trend also. Thus using for example a triploid hybrid of oilseed rape breeding line RfA4 (having Rfk1 gene) and turnip rape breeding line 4021-2 Rfk in a crossing program, could facilitate overall homeologous recombination between A, C and R genome chromosomes. However, Mason et al (2010) found that more complex interactions between genomic structure and alleles are involved controlling homoelogous pairing in *Brassicas*. The irradiation technology has also exploited successfully when developed R2000 *B.napus* Ogu-INRA restorer line (Primard-Brisset et al. 2005). The *Rfo* gene was integrated into C genome of *B.napus* by forcing recombination between radish and rapeseed using ionising irradiation.

Niemelä T, Seppänen M, Badakshi F, Rokka VM, Heslop-Harrison JS. 2012. Size and location of radish chromosome regions carrying the fertility restorer Rfk1 gene in spring turnip rape. Chromosome Research. Apr;20(3):353-361. http://dx.doi.org/10.1007/s10577-012-9280-5 Page 13 of 23.

To have a functional Ogu-INRA CMS/Rf hybrid system for turnip rape, it would be ideal to substitute the putative homeologous representative regions in the A genome with radish chromosome area having the restorer gene. Additional breeding techniques, like changing the ploidy level or using irradiation, to increase the recombination level between R and A genomes may be required. Now the BAC64 as a BAC-FISH probe is reliable for the selection of the turnip rape plants carrying *Rfk1* gene.

## Acknowledgements

We thank for Dr. Regine Delourme for guiding us to get the BAC64 clone for this study. We also thank Xian-Hong Ge for discussions and assistance. The study was supported by National Emergency Supply Agency, Finnish Ministry of Agriculture and Forestry and Foundation of August Johannes and Aino Tiura.

21Wang K, Guo W, Zhang T (2007) Development of one set of chromosome-specific microsatellite-containing BACs and their physical mapping in Gossypium hirsutum L. Theor Appl Genet 115:675-682

Wei W, Li Y, Wang L et al (2010) Development of a novel *Sinapis arvensis* disomic addition line in *Brassica napus* containing the restorer gene for *Nsa* CMS and improved resistance to *Sclerotinia sclerotiorum* and pod shattering. Theor Appl Genet 120:1089-1097

Wijnker E, de Jong H (2008) Managing meiotic recombination in plant breeding. Trends Plant Sci 13:640-646

Xiong Z, Gaeta RT, Pires JC (2011) Homoeologous shuffing and chromosome compensation maintain genome balance in resynthesized allopolyploid *Brassica napus*. PNAS 108:7908-7913

Xiong Z, Pires JC (2010) Karyotype and identification of all homoeologous chromosomes of allopolyploid *Brassica napus* and its diploid progenitors. Genetics 187:37-49
Niemelä T, Seppänen M, Badakshi F, Rokka VM, Heslop-Harrison JS. 2012. Size and location of radish chromosome regions carrying the fertility restorer Rfk1 gene in spring turnip rape. Chromosome Research. Apr;20(3):353-361. http://dx.doi.org/10.1007/s10577-012-9280-5 Page 21 of 23.

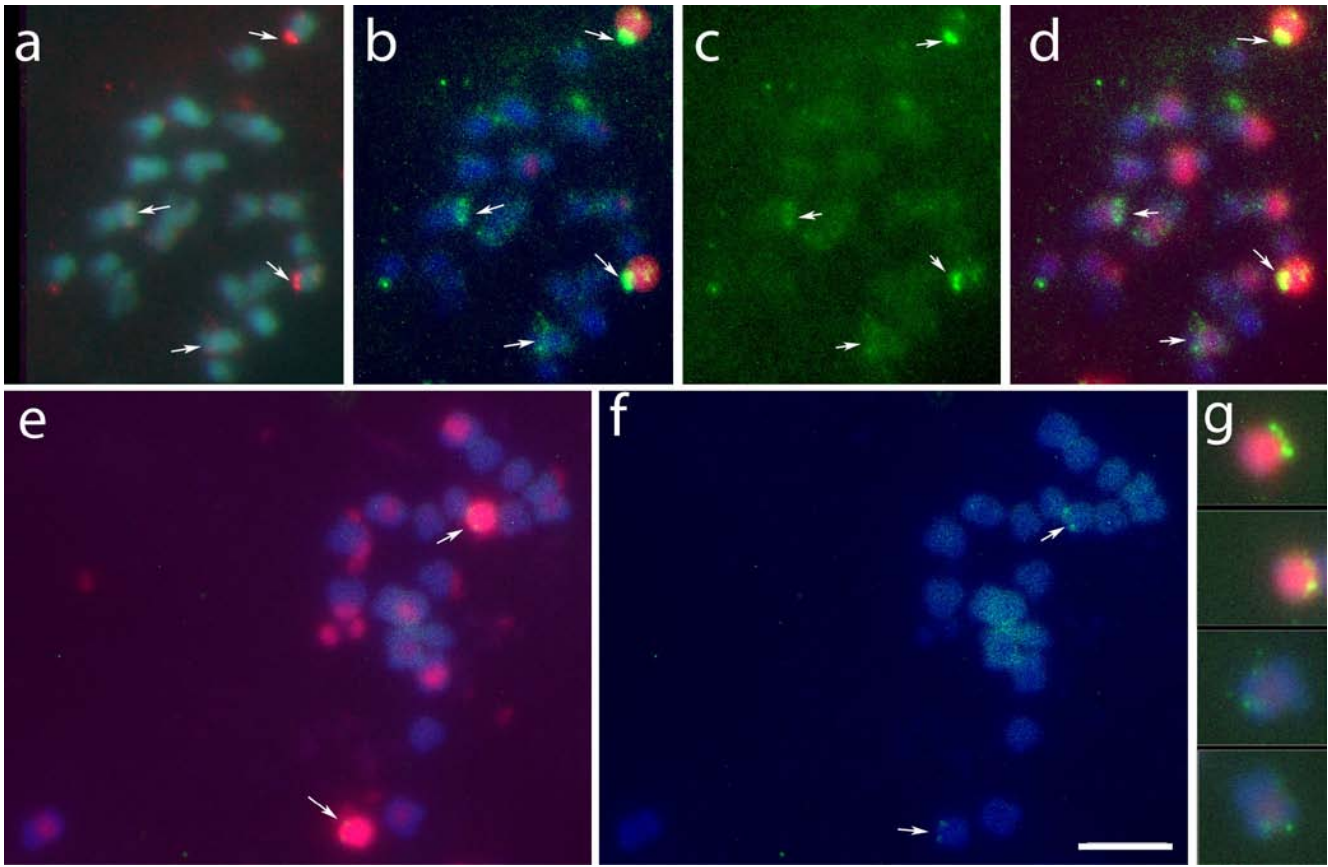

Figure 1. Fluorescence *in situ* hybridization with genomic radish DNA and radish BAC64 clone probe to locate the radish chromosome region carrying fertility restorer *Rfk1* gene in *Brassica rapa* spring turnip rape genome. All the images are from a disomic (2n=2x=20+2R) addition line; chromosomes counterstained blue with DAPI. (a) Double dot signals of BAC64 clone (red) in sub-terminal area on two pairs of sister chromatids; strong signals on additional radish chromosome pair and weaker signals on *B.rapa* chromosome pair (b) Same metaphase as in image (a). BAC64 clone double dot signals green and additional radish chromosome pair labeled red with genomic radish DNA. (c) BAC signal from (b) in green. (d) Image (b) with far red signals shown in red of the 45S rDNA probe. (e) A second metaphase showing the labeled radish chromosome pair red, BAC-FISH signals (green) on the radish-origin chromosomes and far red signals of 45S ribosomal DNA. (f) As (e) showing sub-terminal BAC signal on radish-origin chromosome pair. (g) Isolated chromosomes from a single metaphase showing the BAC64 clone double dot signals (green) in two additional radish-origin chromosomes (red signal) and weaker BAC64 double dots signals (green) on two turnip rape chromosomes. Bar 5µm.

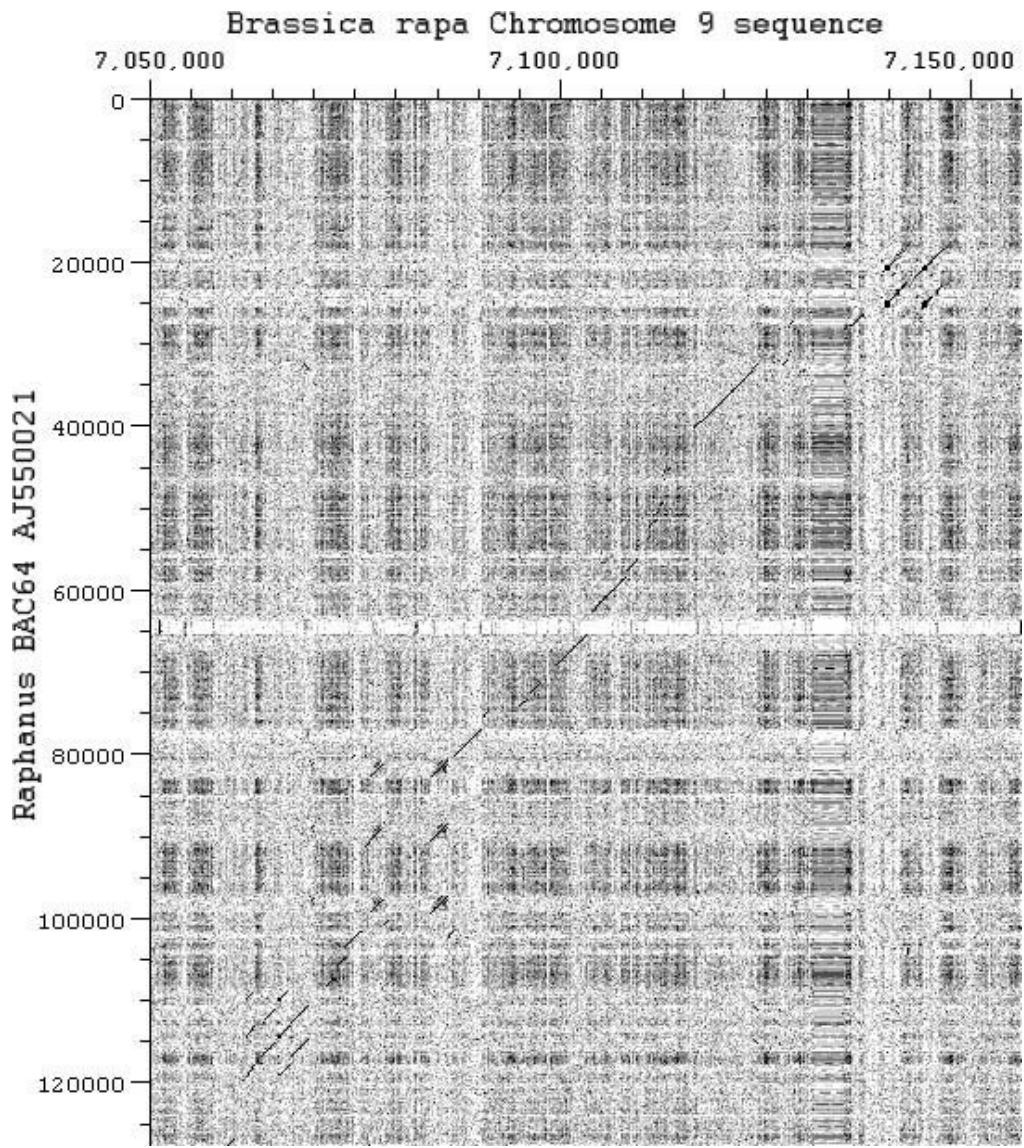

Figure 2. Dot-plot matrix of *Raphanus sativus* BAC64 clone sequence (AJ550021) and *Brassica rapa* chromosome A09 sequence. The sequence from the 125kb long BAC is homologous to a region 7Mb from the end of the 37 Mb long chromosome sequences, consistent with the sub-terminal location identified by *in situ* hybridization.